\documentclass[prl,aps,showpacs,twocolumn,showkeys]{revtex4}
\usepackage{amsmath,amssymb,amsfonts,epsfig,xspace}

\newcommand{\fref}[1]{Figure~\ref{fig:#1}}

\begin{document}
\title{On the Red-Green-Blue Model}
\author{David B. Wilson}
\affiliation{Microsoft Research, One Microsoft Way, Redmond, WA 98052, U.S.A.}
\begin{abstract}
We experimentally study the red-green-blue model, which is a sytem of
loops obtained by superimposing three dimer coverings on offset
hexagonal lattices.  We find that when the boundary conditions are
``flat'', the red-green-blue loops are closely related to SLE$_4$ and
double-dimer loops, which are the loops formed by superimposing two
dimer coverings of the cartesian lattice.  But we also find that the
red-green-blue loops are more tightly nested than the double-dimer loops.
We also investigate the 2D minimum spanning tree, and find that it is
not conformally invariant.
\end{abstract}
\maketitle

\section{Introduction}
We investigate the red-green-blue (RGB) model, which was introduced by
Benjamini and Schramm.  An RGB configuration is a system of the loops
on a region of the triangular lattice, which is obtained by
superimposing three perfect matchings (or dimer coverings) on offset
hexagonal lattices as shown in \fref{rgb}.  The sites of the
triangular lattice may be three-colored so that no two adjacent sites
have the same color.  If we delete the sites of a given color (say
blue), then the sites of the other two colors (red and green) form a
hexagonal lattice on which we can construct a random perfect matching
(the blue perfect matching).  When we superimpose the red, green, and
blue perfect matchings, each vertex is matched with one neighboring
vertex of each of the other two colors.  Since each vertex has degree
two, an RGB configuration consists of closed loops, and each loop has
an orientation if we follow the edges in the order red to green to
blue.

\begin{figure}[tphb]
\begin{center}
\epsfig{figure=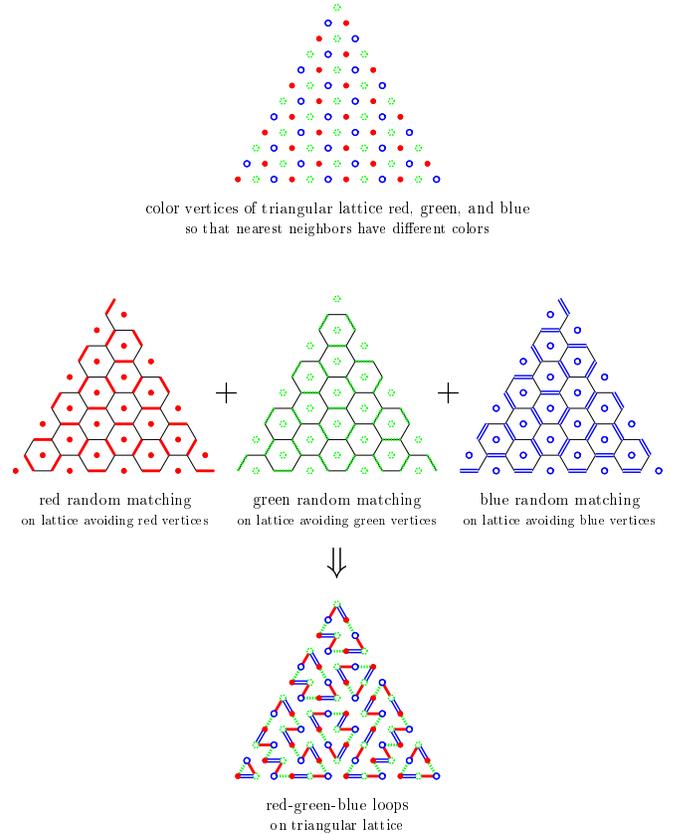,width=\columnwidth}
\caption{The red-green-blue (RGB) model on a triangular region of side length $12$ ($L=4$).  The three color classes of vertices are balanced along the boundary, so the boundary conditions are ``flat''.}
\label{fig:rgb}
\vspace*{-4pt}
\end{center}
\end{figure}

It is worth remarking that the boundary conditions of dimer systems
can have a profound impact on the behavior of dimers even far
within the interior of a region \cite{cohn-elkies-propp:arctan,cohn-larsen-propp:hexagon,cohn-kenyon-propp}.  There are height functions
associated with dimer configurations on the cartesian lattice
\cite{levitov:equivalence,zheng-sachdev:quantum} and
hexagonal lattice \cite{blote-hilhorst:roughening}
(see also \cite{thurston:conway,propp:tilings}).  If there is an imbalance
between the different colors of vertices along the boundary, then
the height along the boundary will be ``tilted'', and this affects the
dimers throughout the region.  Consequently, the behavior of the RGB
model on regions with tilted boundary conditions could be
different from the behavior of the RGB model on the regions that we
consider here, where the three
color classes along the boundary are balanced (``flat'' boundary conditions).

In earlier work, Kenyon and the author \cite{kenyon-wilson:rgb} found
experimentally that the fractal dimension of these loops is $3/2$.
Here we report on additional experiments, where we find that the
winding angle variance at a typical point on a loop is $1\times \log
D$ where $D$ is the diameter of the loop, and that the system of loops
appears to be conformally invariant.  These properties suggest that
the RGB loops are closely related to stochastic Loewner
evolution \cite{S} with parameter $\kappa=4$ (SLE$_4$),
and that the RGB loops belong to the same universality class as the contours of 2D
Fortuin-Kasteleyn \cite{FK} clusters at criticality when $q=4$, and
the loops formed in the double-dimer (or ``double-domino'') model
\cite{kondev-henley}, which in turn are thought to correspond to the
``contours'' of a Gaussian free field \cite{kondev-henley}.
However, the system of
RGB loops does not have the same limiting behavior as the system of
double-dimer loops, because we also find that the RGB
loops are more tightly nested than the double-dimer loops.

\begin{figure*}[bhpt]
\begin{center}
\epsfig{figure=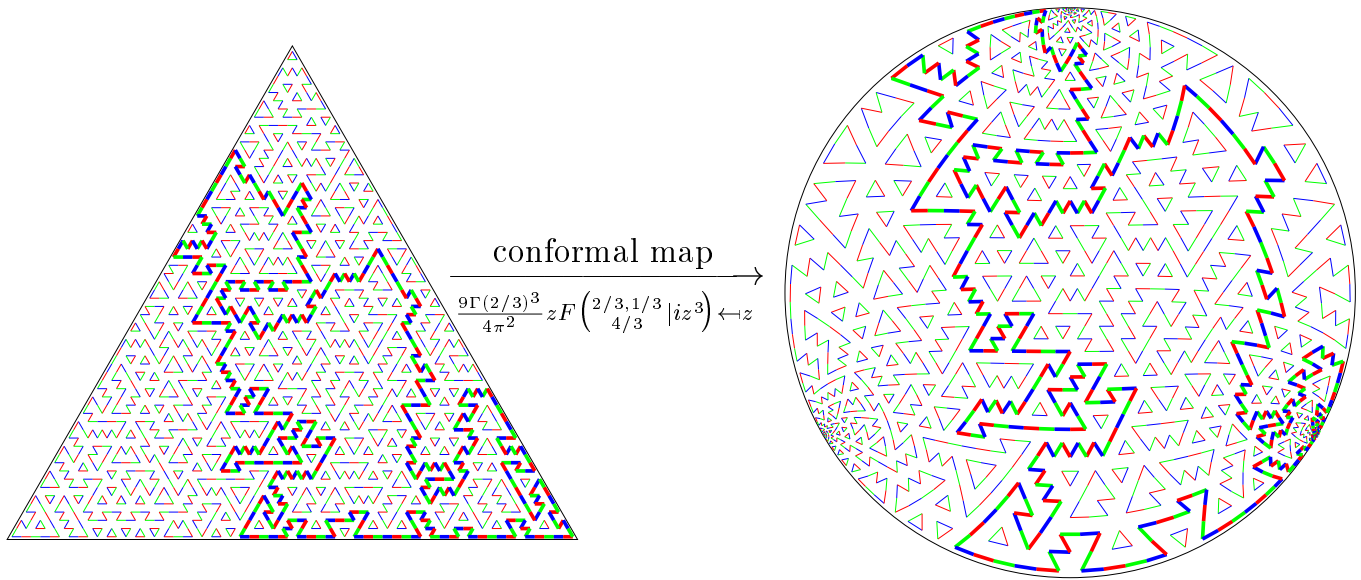,width=\textwidth}
\caption{The RGB model on the triangular domain ($L=20$) is conformally mapped to a disk.  The inverse conformal map, from the disk to the triangle, is the hypergeometric function shown in the diagram.  If the RGB model is conformally invariant, then the loops in disk will be rotationally invariant.}
\label{fig:conf}
\end{center}
\end{figure*}

\section{Generating RGB configurations}
To generate an RGB configuration of a region, we need to generate
dimer coverings of three regions of the hexagonal lattice.  There are
many ways to generate dimer coverings of the the hexagonal lattice,
but the fastest of these is based on a generalization of Temperley's
bijection \cite{T} between spanning trees and dimers.  To generate the perfect
matchings on the hexagonal lattice, the corresponding spanning trees
are on a directed triangular lattice (see
\cite{kenyon-propp-wilson:temperley} for details), and these spanning
trees may be quickly generated using an algorithm based on loop-erased
random walk \cite{PW2}.  This
generalized-Temperley bijection only works for regions of the
hexagonal lattice that have certain special (very flat) boundary
conditions, so we can only expect to use it to generate RGB
configurations of certain nice regions.  One region where we can use
spanning trees to rapidly generate RGB configurations is the
equilateral triangle with side length $3 L$ (Figures~\ref{fig:rgb} \& \ref{fig:conf}).


\section{Windiness of RGB loops}
We recall the definition of the windiness of a loop used in
\cite{wieland-wilson:winding}.  Consider an ant which travels along
the loop; after the ant has just traversed a
given edge in the loop, before traversing the next edge it will either
turn left $120^\circ$, turn right $120^\circ$, or not turn at all.  If
we keep track of the total turning (measured in radians) when the ant
travels from point $A$ to point $B$ on the loop, then this is
(approximately) the winding angle between points $A$ and $B$. When the ant
travels all the way around the loop, it has turned $\pm360^\circ$, so to
make the winding angle between points $A$ and $B$ independent of the
number of times that the ant travels around the loop and the direction
of travel, we adjust the total turning (measured in radians) by $2\pi
\times [\text{\# steps between $A$ and $B$}] \div [\text{length of
loop}]$.
To define the winding angle at a given point $X$ relative to the global
average direction, we pick an arbitrary point $A$, compute the winding
angle from $A$ to $X$, and subtract a global constant so that the average
winding angle at points on the curve is $0$.

When we measure the variance in the winding angle at random points
along the longest loop in the RGB configuration in a region of order
$L$, we find that the variance grows like $1\times\log L$ --- so in
the notation of \cite{wieland-wilson:winding}, $\kappa_2=1$.  This
winding angle variance coefficient of 1 also shows up in in the
contours of FK clusters at criticality when $q=4$, and other related
models such as the double-dimer model, and SLE$_4$ (see
\cite{duplantier-saleur:winding,wieland-wilson:winding,duplantier-binder:winding}).

\section{Conformal invariance of RGB loops}

Since there is only one region, the equilateral triangle, for which we
can \textit{rapidly\/} generate RGB configurations, this makes the
testing of conformal invariance somewhat interesting.  The test that
we use is similar in spirit to the tests used by Schramm to test the
conformal invariance of the uniform spanning tree.  We conformally
mapped the RGB model on the triangular domain to a circular domain, as
shown in \fref{conf}.
  If the RGB model with flat boundary conditions
were conformally invariant, then it must be that after we map a region
to the disk, the resulting system of loops would be rotationally
invariant.  But if conformal invariance failed to hold, then there
would be no particular reason to believe that the loops mapped to the
disk would be rotationally invariant.  After all, referring to
\fref{conf}, the points in the disk to which the corners of the
triangle are mapped certainly look different than other points in the
disk, so \textit{a priori\/} we would expect the image in the disk to
be anisotropic if the RGB model were not conformally invariant.
As we shall see, the minimum spanning tree model fails this test,
so this test is a nontrivial test of conformal invariance.

To test the rotational invariance of the image of the RGB model in the
disk, we singled out the outermost loop surrounding the center of
the circular domain, and collected statistics on its furthest extents
in the $\pm x$- and $\pm y$-directions.  If the loops in the RGB model
are conformally invariant, then these four random variables would be
equidistributed.  But otherwise, there would be no particular reason
to believe that any of these random variables (other than the first
two) would have the same distributions.  As shown in \fref{rgb-cdf},
the cumulative distribution functions for these four random variables
appear to coincide, so we conclude that the RGB model appears to be
conformally invariant.

\begin{figure}[hbtp]
\begin{center}
\epsfig{figure=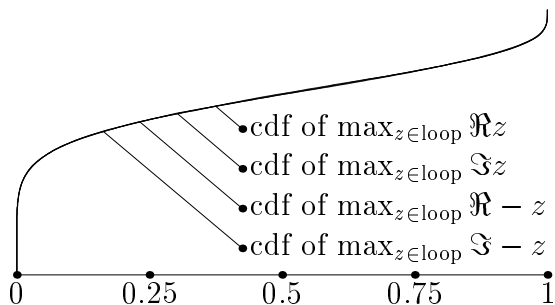}
\end{center}
\caption{In the RGB model ($L=1024$), the outermost loop surrounding the center point is selected, and conformally mapped to the disk.  The image of this loop appears to be rotationally invariant, in particular, the empirical cumulative distribution functions for furthest extents of the loop in the $x$-direction, $y$-direction, $-x$-direction, and $-y$-direction appear to coincide.}
\label{fig:rgb-cdf}
\end{figure}

\section{Conformal \textit{non\/}-invariance of minimum spanning trees}

To evaluate the efficacy of our conformal invariance test, we applied
it to two additional models: the minimum spanning tree and uniform
spanning tree models.  The minimum spanning tree (MST) is formed by
assigning uniformly random edge weights to the edges of the cartesian
lattice, and picking the spanning tree (connected acyclic subset of
\begin{figure}[tphb]
\begin{center}
\epsfig{figure=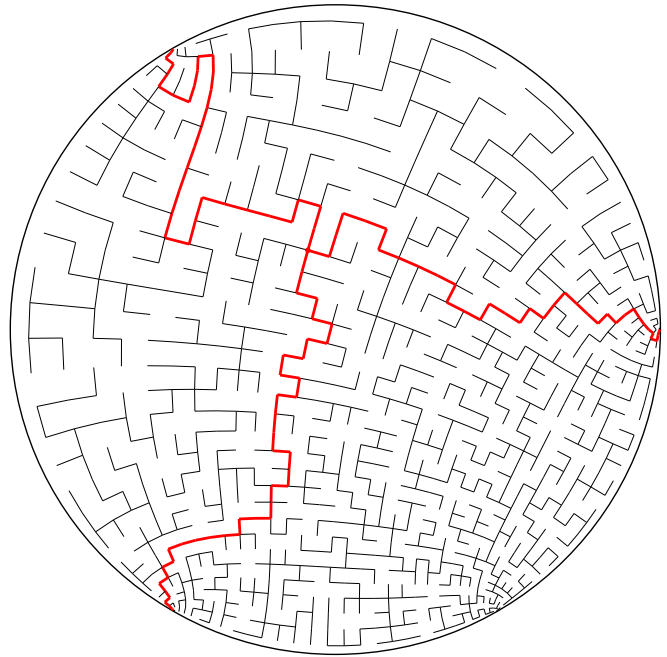,width=\columnwidth,angle=-90}
\end{center}
\caption{The minimum spanning tree (MST) on the $32\times 32$ square grid, mapped conformally to the disk so that three of the corners of the square are mapped to the cube roots of unity.  The paths connecting these corners is highlighted; the triple point $T$ is the point contained in all three paths.}
\label{fig:mst-disk}
\end{figure}
edges) which minimizes the total weight.  The uniform spanning tree
(UST) is simply a spanning tree chosen uniformly at random from all
spanning trees.

\fref{mst-disk} shows the MST of a square grid after it is conformally mapped
to the unit disk, with three of the corners mapped to the three cube
roots of unity.  To test the rotational invariance of the MST after it
is mapped to the disk, we looked at the paths connecting the three
points at the cube roots of unity (highlighted in \fref{mst-disk}),
and focused on the ``triple-point'' $T$ contained in all three paths.
If the image of MST in the disk were isotropic, then $T$, $e^{2\pi i/3}
T$, and $e^{4\pi i/3} T$ would be equidistributed.  However, as
\fref{mst-cdf} illustrates, these variables are not equidistributed,
so we conclude that the MST is not conformally invariant.
The conformal non-invariance of the MST is surprising, given
the close relationship between the MST and invasion percolation \cite{CCN},
the close relationship between invasion percolation and percolation,
and the conformal invariance of percolation \cite{LPS,Car92,Smirnov}.

\begin{figure}[tphb]
\begin{center}
\epsfig{figure=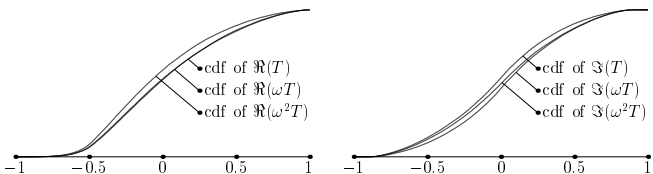,width=\columnwidth}
\end{center}
\caption{The distribution of the triple point $T$ for the MST (on the $1024\times 1024$ grid) is not symmetric under $120^\circ$ rotations, as it would be if the MST were conformally invariant.  ($\omega=\exp(2\pi i/3)$.)}
\label{fig:mst-cdf}
\end{figure}

In contrast to the MST, the UST passes this test, as shown in
\fref{ust-cdf}.  The triple point $T$ connecting three boundary points
of the UST is already known to be conformally invariant \cite{K2}, and
indeed the entire UST process is now known to be conformally invariant
\cite{LSW:2-8}.

\begin{figure}[tphb]
\begin{center}
\epsfig{figure=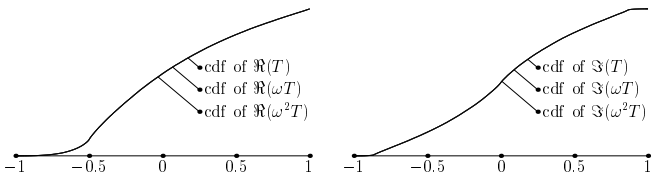,width=\columnwidth}
\end{center}
\caption{In contrast, the distribution of the triple point $T$ for the UST (on the $1024\times 1024$ grid) passes this conformal invariance test.  Kenyon \cite{K2} has proven that the triple point for UST is in fact conformally invariant (see also \cite{LSW:2-8}).}
\label{fig:ust-cdf}
\end{figure}

Thus we learn not only that the minimum spanning tree is not
conformally invariant, but that this test is a nontrivial test of
conformal invariance.

\section{Nesting of RGB loops}

For a scale-invariant loop model on a region of side length $L$, we
would expect the number of loops surrounding a point to scale as
$\text{const.}\times \log L$.  The value of this constant is a measure
of how deeply nested the loops are.  For the double-dimer model,
Kenyon \cite{kenyon:dimers} proved that this nesting constant is
$1/\pi^2$.  We measured the nesting constant of the RGB loops, and
found that it is $20\%$--$25\%$ larger than the double-dimer nesting
constant, but we
do not have a guess for its exact value.  \fref{rgbdd} shows that the
outermost red-green-blue loops are in a sense larger than the
outermost double-dimer loops, which is consistent with the
red-green-blue loops being more tightly nested within one another.

\begin{figure}[tphb]
\begin{center}
\epsfig{figure=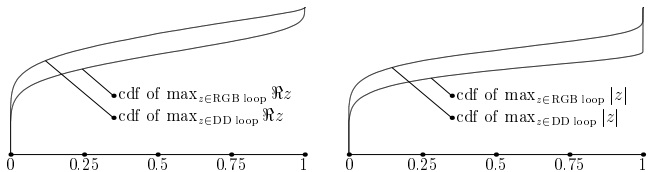,width=\columnwidth}
\end{center}
\caption{Comparison of the outermost loop surrounding in the origin within the red-green-blue and double-dimer models.  The cdf for the size of the RGB loop is smaller than the corresponding CDF for the DD loop, so in this sense the outermost RGB loop is larger than the outermost double-dimer loop.  There is a good chance that the outermost RGB loop approaches the boundary quite closely, as in \fref{conf}.}
\label{fig:rgbdd}
\end{figure}

\section{Conclusions}

Our experiments indicate that the loops of the RGB model (with flat
boundary conditions) are conformally invariant and have windiness
constant $1$.  Earlier experiments \cite{kenyon-wilson:rgb} have
indicated that the fractal dimension is $3/2$.  These properties
suggest that RGB loops belong to the same universality class as
double-dimer loops, the fully-packed-loop model with $n=2$, and the
contours of critical FK clusters with $q=4$, and are closely related
to SLE$_4$.  However, the \textit{system\/} of RGB loops (not just
individual loops) differs from the system of double-dimer loops,
because the loops are nested within one another more tightly.

\section*{Acknowledgements}
We thank Oded Schramm for useful discussions.

\bibliography{rgb,ww}

\end{document}